\documentclass[12pt,a4paper]{article}
\usepackage[T2A]{fontenc}
\usepackage[utf8]{inputenc}
\usepackage[english]{babel}
\usepackage{amssymb}
\usepackage{amsmath}
\usepackage{graphicx}
\usepackage[small,sc,center]{titlesec}
\usepackage{indentfirst}
\usepackage{siunitx}
\usepackage[labelsep=period]{caption}
\usepackage{caption}

\newcommand{\nc}{\newcommand}
\nc{\etab}{\eta_\mathrm{B}}
\nc{\Menv}{M_\mathrm{env}}
\nc{\Teff}{T_\mathrm{eff}}
\nc{\tev}{t_\mathrm{ev}}

\addto\captionsenglish{}

\begin{document}

\begin{center}
	\textbf{Evolutionary status of W Vir pulsating variables}
	
	\vskip 3mm
	\textbf{Yu. A. Fadeyev\footnote{E--mail: fadeyev@inasan.ru}}
	
	\textit{Institute of Astronomy, Russian Academy of Sciences,
		Pyatnitskaya ul. 48, Moscow, 119017 Russia} \\
	
	Received October 12, 2020; revised October 16, 2020; accepted October 27, 2020
\end{center}

\textbf{Abstract} ---
Stellar evolution calculations for population II stars with initial composition $Y_0=0.25$,
$Z_0=10^{-3}$ and the initial stellar mass $M_0 = 0.82M_\odot$ were carried out from the main
sequence to the white dwarf stage.
Twelve AGB and post--AGB evolutionary sequences were computed with different values of the
parameter in the Bl\"ocker mass loss rate formula ($0.01\le\etab\le 0.12$).
Selected models of evolutionary sequences with masses $M=0.536M_\odot$, $0.530M_\odot$ and
$0.526M_\odot$ that experience the loop in the Hertzsprung--Russel diagram due to the final
helium flash were used as initial conditions for solution of the equations of hydrodynamics
describing radial stellar oscillations.
The region of instability to radial fundamental mode pulsations is shown to extend from
the asymptotic giant branch to effective temperatures as high as
$\Teff\approx 6\times 10^3$ K.
Pulsation periods of hydrodynamic models are in the range from 15 to 50 day and agree with
periods of W~Vir pulsating stars.
The models of intermediate spectral type fundamental mode pulsators with periods $\Pi > 50$
day locate in the upper part of the Hertzsprung--Russel diagram in the region of semiregular
pulsating variables.
We conclude that W~Vir pulsating variables are the low--mass post--AGB stars that
experience the final helium flash.


\newpage
\section*{introduction}

W Vir pulsating variables belong to population~II cepheids and are observed among field
stars as well as in globular clusters.
Periods of these variables range from 8 to 35 day (Samus' et al. 2017).
W~Vir stars are remarkable due to strong hydrogen emission lines observed during rising light
(Joy 1937).
Doubled metallic absorption lines detected simultaneously with hydrogen emission are
explained by the model of periodic shock wave propagating through outer expanding layers
of the stellar atmosphere (Abt 1954; Whitney 1956; Wallerstein 1959).
Periodic shock waves in the stellar atmosphere are due to nonlinear effects
in the large amplitude radial oscillations.
Stronger hydrogen emission is observed in RV~Tau stars with higher luminosities and
longer periods.
RV Tau semiregular pulsating variables are believed to be connected with W~Vir stars
(Wallerstein 2002).

W~Vir variability is thought to appear in low--metallicity low--mass double shell source
stars (Wallerstein and Cox 1984).
Stellar evolution calculations by Schwarzschild and H\"arm (1970) showed that thermal
instability of the helium burning shell may lead to occurence of the loop in the evolutionary
track so that the star crosses the cepheid instability strip in the Hertzsprung--Russel (HR)
diagram.
More detailed evolutionary calculations by Gingold (1974, 1976) and Sweigart et al. (1976)
revealed that the necessary condition for occurence of the blue loop in the HR diagram is that
the helium flash occurs at the onset of the post--AGB phase when the hydrogen envelope
mass becomes as small as $\Menv\approx 0.02M$, where $M$ is the mass of the star.

All studies of the evolutionary status of W~Vir stars were done on the basis of stellar
evolution calculations whereas occurence of radial pulsations was supposed from
consideration of the loop on the evolutionary track with respect to the instability strip
of Cepheids and RR Lyr variables.
For example, Gingold (1976) extrapolated results of RR Lyr linear model calculations by
Tuggle and Iben (1972) to higher luminosities.
On the other hand, computations of nonlinear pulsation models of W~Vir stars
(Fadeyev and Fokin 1985; Buchler and Kovacs 1987, Kovacs and Buchler 1988;
Fadeyev and Muthsam 1990; Serre et al. 1996; Smolec 2016)
were carried out without a direct relation between the stellar evolution models and
the initial conditions of radiation hydrodynamics calculations.
Below we show that pulsations of W~Vir stars strongly depend on the mass and luminosity
because of the low gas density in the envelope of the pulsating star and significant
contribution of radiation energy density to the internal energy of stellar matter.
For example, variation of the stellar mass within 4\% is accompanied by change of the luminosity
by a factor of two.

The goal of the present study is to establish evolutionary status of W~Vir pulsating variables
on the basis of consistent stellar evolution and nonlinear stellar pulsation calculations.
Methods of the solution are discussed in our earlier papers devoted to studies of classical
Cepheids (Fadeyev 2018) and post--AGB stars (Fadeyev 2019a, 2019b).

\section*{results of stellar evolution calculations}

In the present work we consider stellar evolution from the main sequence to the stage
of a white dwarf with luminosity $L\sim L_\odot$.
Initial mass--fraction abundances of helium and heavier elements were assumed to agree
with those of typical globular clusters: $Y_0=0.25$ (Salaris et al. 2004)
and $Z_0=0.001$ (Catelan 2004).
The initial mass of the star on the zero age main sequence is $M_0=0.82 M_\odot$
so that the age of W~Vir models is $\tev = 1.244\times 10^{10}$~yr.

Evolutionary sequences were computed with the program MESA version 12778 (Paxton et al. 2019).
In order to calculate the thermonuclear energy release and evolution of isotope abyndances
we employed the reaction network 'pp\_cno\_extras\_o18\_ne22.net' with 26 nuclei
from hydrogen ${}^1\mathrm{H}$ to magnesium ${}^{24}\mathrm{Mg}$ coupled by 81 reactions.
The reaction rates were calculated with the JINA Reaclib data (Cyburt et al. 2010).
Convective mixing was treated according to the theory of B\"ohm--Vitense (1958) for the
mixing length to pressure scale height ratio $\alpha_\mathrm{MLT} = \Lambda/H_\mathrm{P} = 2.0$.
Mixing due to overshooting at the boundaries of convective stability was computed according
to Herwig (2000) with parameters $f=0.016$ and $f_0=0.004$.
Evolutionary calculations of the core helium burning stage with option 'conv\_premix\_avoid\_increase'
allowed us to significantly diminish the role of breathing pulses and to avoid appearence
of spurious loops on the evolutionary track.
The mass loss rate at evolutionary stages preceding AGB was evaluated by the Reimers (1975)
formula with parameter $\eta_\mathrm{R} = 0.5$, whereas at later stages we used the mass
loss rate formula  by Bl\"ocker (1995).
Bearing in mind existing uncertainties in mass loss rates of AGB stars we computed twelve
AGB and post--AGB evolutionary sequences with parameter $\etab$ equidistantly distributed
in the range $0.01\le\etab\le 0.12$.

Results of stellar evolution calculations are illustrated in Fig.~\ref{fig1}, where three tracks
of AGB and post--AGB stages are shown in the HR diagram for the mass loss parameters
$\etab = 0.03$, 0.07 and 0.1.
According to Miller Bertolami (2016) we assume that the post--AGB stage begins when
the mass of the hydrogen envelope is $\Menv \le 0.01M$.
Transition from the AGB to post--AGB stage is marked on each track by the open circle,
whereas the filled circle corresponds to the peak of the helium--flash luminosity.

Of most interest for the present study are the evolutionary changes of the star
in the effective temperature range
 $4\times 10^3\:\mathrm{K}\lesssim\Teff\lesssim 6\times 10^3\:\mathrm{K}$.
Evolutionary tracks within this effective temperature interval are shown
with enlarged scale in Fig.~\ref{fig2}.

Main properties of evolutionary sequences are listed in Table~\ref{tabl1}, where
$n_\mathrm{TP}$ and $\langle\Delta t_{ip}\rangle$ are the number of thermal pulses during
the AGB stage and the mean time interval between two thermal pulses,
$M$ and $L$ are the mass and luminosity of the star at the onset of the post--AGB stage
when the mass of the hydrogen envelope is $\Menv=0.01M$,
$\Menv^\star$ is the mass of the hydrogen envelope at the peak of the helium luminosity
during the final helium flash.
The sign '$-$' in the last column implies absence of the blue loop due to the final
helium flash in the HR diagram.

Increase of the mass loss rate leads to the shorter AGB lifetime and to the smaller
number of thermal pulses $n_\mathrm{TP}$.
Two temporal dependences of the helium--burning shell luminosity $L_\mathrm{He}$ for
evolutionary sequences $\etab=0.01$ и $\etab=0.1$ are shown in Fig.~\ref{fig3}, where
for the sake of graphical representation the evolutionary time $\tev$ is set to zero at
the helium peak luminosity of the first thermal pulse $t_{\mathrm{TP}1}$.
As seen from these plots, the TP--AGB lifetime reduces by nearly two times when the mass
loss rate increases by ten times although the mean time interval between flashes remains
almost unchanged.

The necessary condition for occurence of the loop in the evolutionary track during
the post--AGB stage is that the final thermal pulse occurs at the mass of the hydrogen
envelope $\Menv\approx 0.02M$ (Gingold 1974).
As seen in Table 1, results of our computations agree with this condition because
in the three evolutionary sequences without the loop on the evolutionary track
($\etab=0.05$, 0.06 and 0.12)
the mass of the hydrogen envelope during the final thermal pulse is significantly larger.
It should be noted that for $\etab=0.08$ the mass of the hydrogen envelope at the final
thermal pulse is $\Menv=0.020M$ but the loop of the evolutionary track extends in the HR
diagram to effective temperatures as high as $\Teff\approx 5000$ K.
That is why in Table~\ref{tabl1} this evolutionary sequence is marked as that without
the loop.
This example allows us to conclude that the sufficiently small value of the mass of
the hydrogen envelope during the final helium flash is not sufficient for occurence
of the extended loop on the evolutionary track.

\section*{results of stellar pulsation calculations}

In consistent stellar evolution and nonlinear stellar pulsation calculations selected
evolutionary models are used as initial conditions for solution of the equations of
radiation hydrodynamics.
In the present work the system of the equations of hydrodynamics is extended due to
transport equations of time--dependent turbulent convection (Kuhfu\ss\ 1986) and is
discussed in our earlier paper (Fadeyev 2013).

Solution of the Cauchy problem for the equations of hydrodynamics implies that the role
of initial perturbations is played by interpolation errors that appear when the
physical variables of the evolutionary model (the radius, the luminosity, the temperature
and the pressure) are used for calculation in the Lagrangian grid of the
hydrodynamic model.
Evolutionary calculations of the final AGB stage were done with the number of mass
zones $N_\mathrm{MESA}\sim 10^4$, whereas the number of mass zones of the hydrodynamic
models was $N=500$.
The inner boundary of the hydrodynamic model was determined in the layers with
radiative transfer and the gas temperature $T\sim 10^6$ K.
The radius of the inner boundary is $r\lesssim 0.05R$, where $R$ is the radius of the
evolutionary model.
The radius and the luminosity of the inner boundary were assumed to be constant
on the time interval of the Cauchy problem solution.

The solution of the equations of hydrodynamics describes decaying oscillations if the star
model is stable to radial pulsations but in the case of the pulsational instability
the solution describes oscillations with exponentially increasing amplitude
until the limiting amplitude is reached.
The main difference between classical Cepheids and W~Vir pulsating variables is that
the stars of the latter group have significantly higher rates of the
amplitude growth.
For example, typical growth rate of the kinetic energy of pulsation motions in W~Vir
models is $\eta = \Pi d\ln E_\mathrm{K,max}/dt\approx 1$, where $\Pi$ is the pulsation period
and $E_\mathrm{K,max}$ is the maximum kinetic energy reached during the period.

\begin{figure}
\centerline{\includegraphics{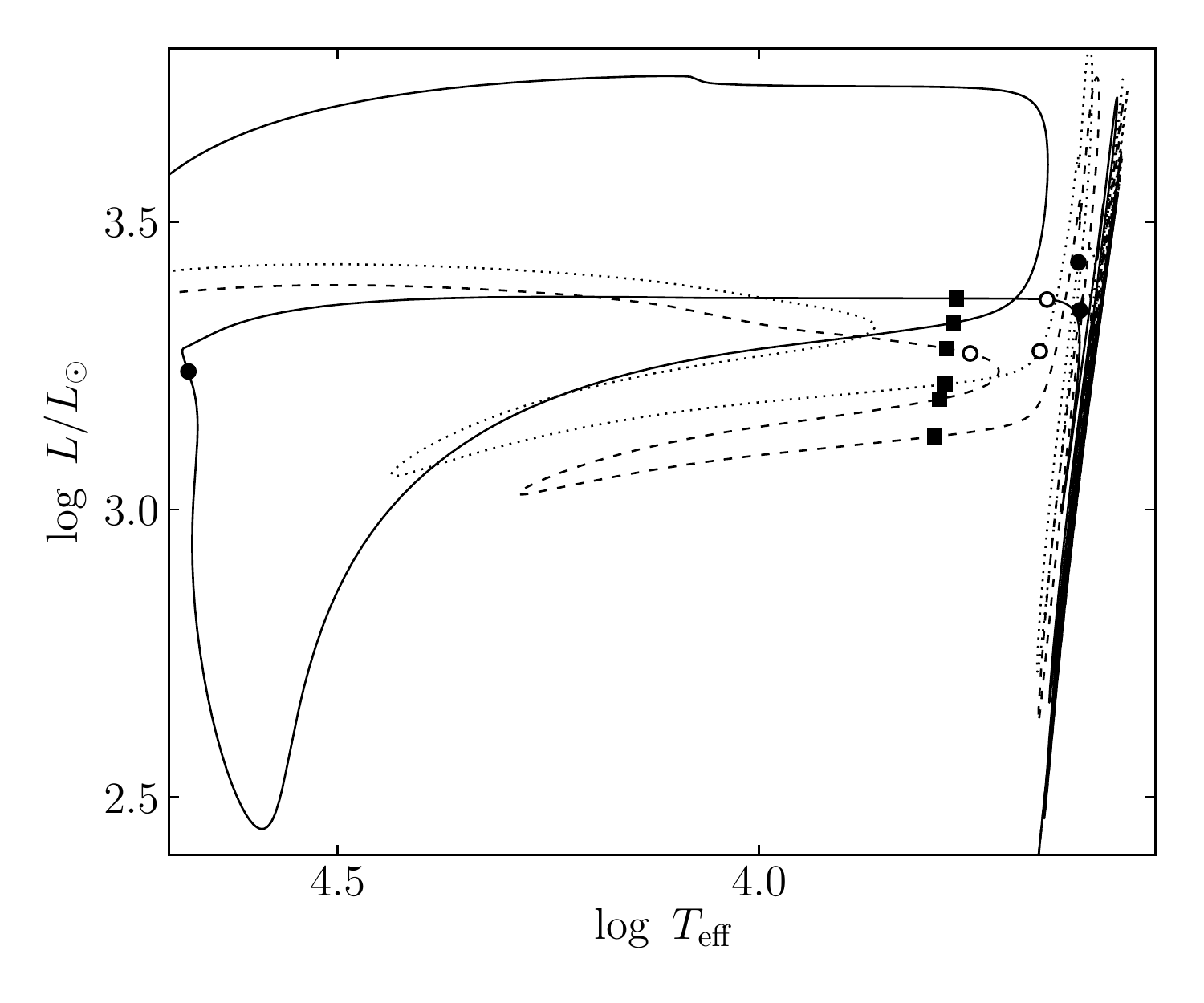}}
\caption{Evolutionary tracks in the HR diagram for the mass loss parameters
    $\etab=0.1$ (solid line), $\etab=0.07$ (dashed line) and $\etab=0.03$ (dotted line).
    Filled circles mark maxima of the helium burning shell during the final helium flash,
    open circles mark the onset of the post--AGB stage, filled squares mark the blue edge
    of the pulsation instability region determined from hydrodynamic computations.}
\label{fig1}
\end{figure}

\begin{figure}
\centerline{\includegraphics{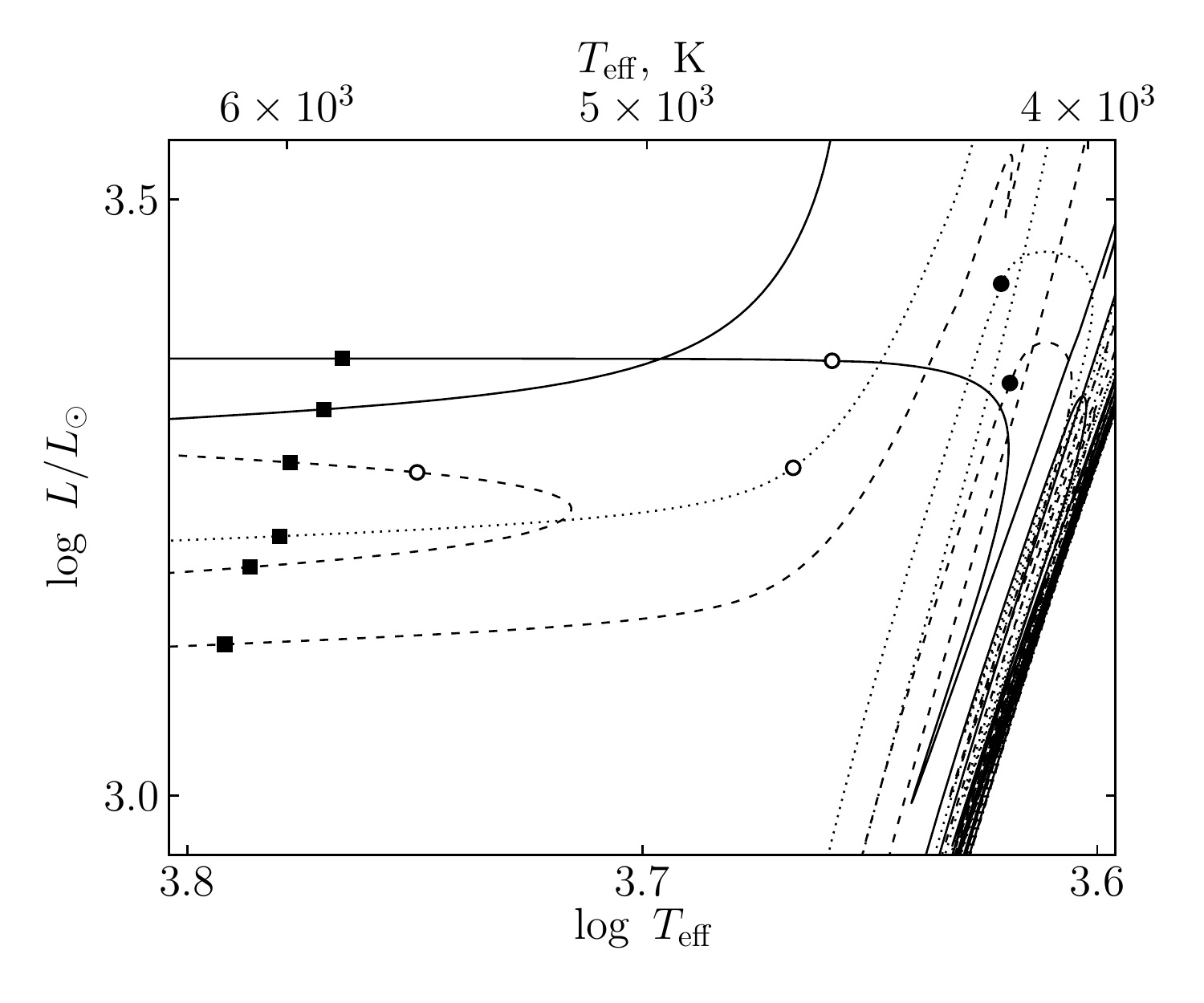}}
\caption{Same as Fig.~\ref{fig1} but with enlarged scales of the HR diagram area
    confined by the asymptotic giant branch and the blue edge of the pulsation instability
    region.}
\label{fig2}
\end{figure}

\begin{figure}
\centerline{\includegraphics{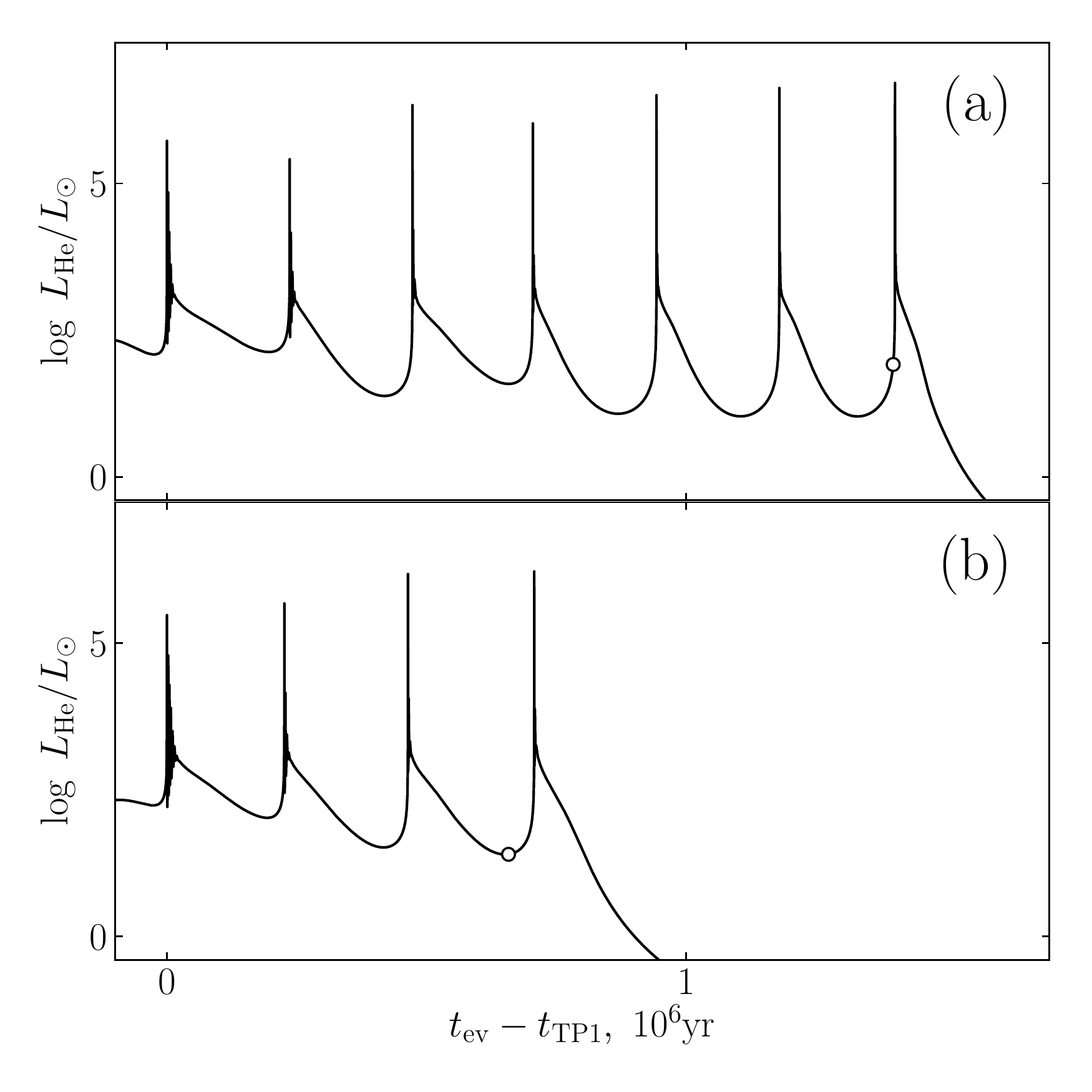}}
\caption{The helium--burning shell luminosity $L_\mathrm{He}$ during the
    AGB stage for the mass loss rate parameters $\etab=0.01$ (a) and $\etab=0.1$ (b).
    Open circles on the plots mark the onset of the post--AGB stage when $\Menv=0.01M$.
    The evolutionary time $\tev$ on the horizontal axis is set to zero at the helium peak
    luminosity of the first thermal pulse $t_{\mathrm{TP}1}$.}
\label{fig3}
\end{figure}

Hydrodynamic calculations of nonlinear stellar pulsations were carried out for selected models
of evolutionary sequences computed with mass loss rate parameters $\etab = 0.03$, 0.07 and 0.1.
Instability to radial fundamental mode pulsations was found for stars between AGB and
effective temperatures as high as $\Teff\approx 6000$ K.
Estimates of $\Teff$ corresponding to the blue edge of pulsational instability ($\eta=0$) are
shown by filled squares in Figs.~\ref{fig1} and \ref{fig2}.

In order to establish the evolutionary status of W~Vir stars we have to compare the
pulsation periods of hydrodynamic models with periods of observed light variations.
As seen in Fig.~\ref{fig1}, for the parameter $\etab=0.03$ the evolutionary track crosses
the instability edge only once during the evolution from AGB.
The extension of the red loop in the HR diagram is not sufficient and the track
turns back at the effective temperature $\Teff \approx 7300$ K.
Hydrodynamic models of evolutionary sequences $\etab=0.07$ and $\etab=0.1$ appear in the
region of pulsational instability two times: during evolution from AGB and when the
evolutionary track loops to the red giant region in the HR diagram and the effective
temperature becomes $\Teff < 6000$~K.

\begin{figure}
\centerline{\includegraphics{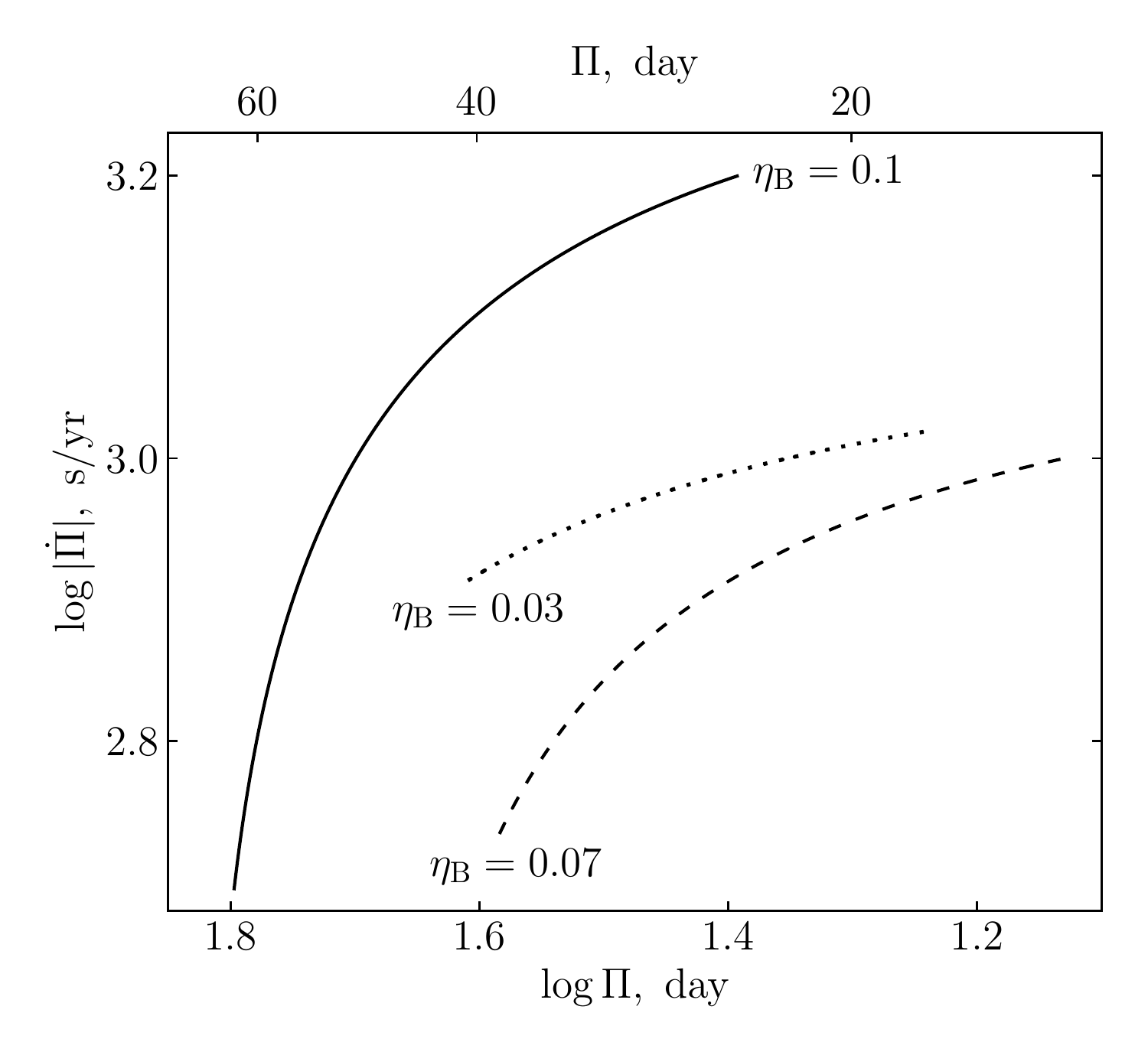}}
\caption{The rate of period change $\dot\Pi$ as a function of period $\Pi$
    for the evolutionary stage from AGB up to the first crossing of the blue edge of the
    pulsation instability region.}
\label{fig4}
\end{figure}

The plots of the period change rate as a function of period for the stage of evolution
from AGB to the first crossing of the blue edge of the instability region are shown
in Fig.~\ref{fig4}.
The initial point of each plot represents the hydrodynamic model with effective temperature
$\Teff\approx 5000$~K ($\etab=0.03$, 0.07) and $\Teff\approx 4500$~K ($\etab=0.1$).
Evolution proceeds with decreasing period from left to right to the blue edge of the pulsation
instability region and lasts from 2200 yr for $\etab=0.03$ to 3200 yr for $\etab=0.1$.

\begin{figure}
\centerline{\includegraphics{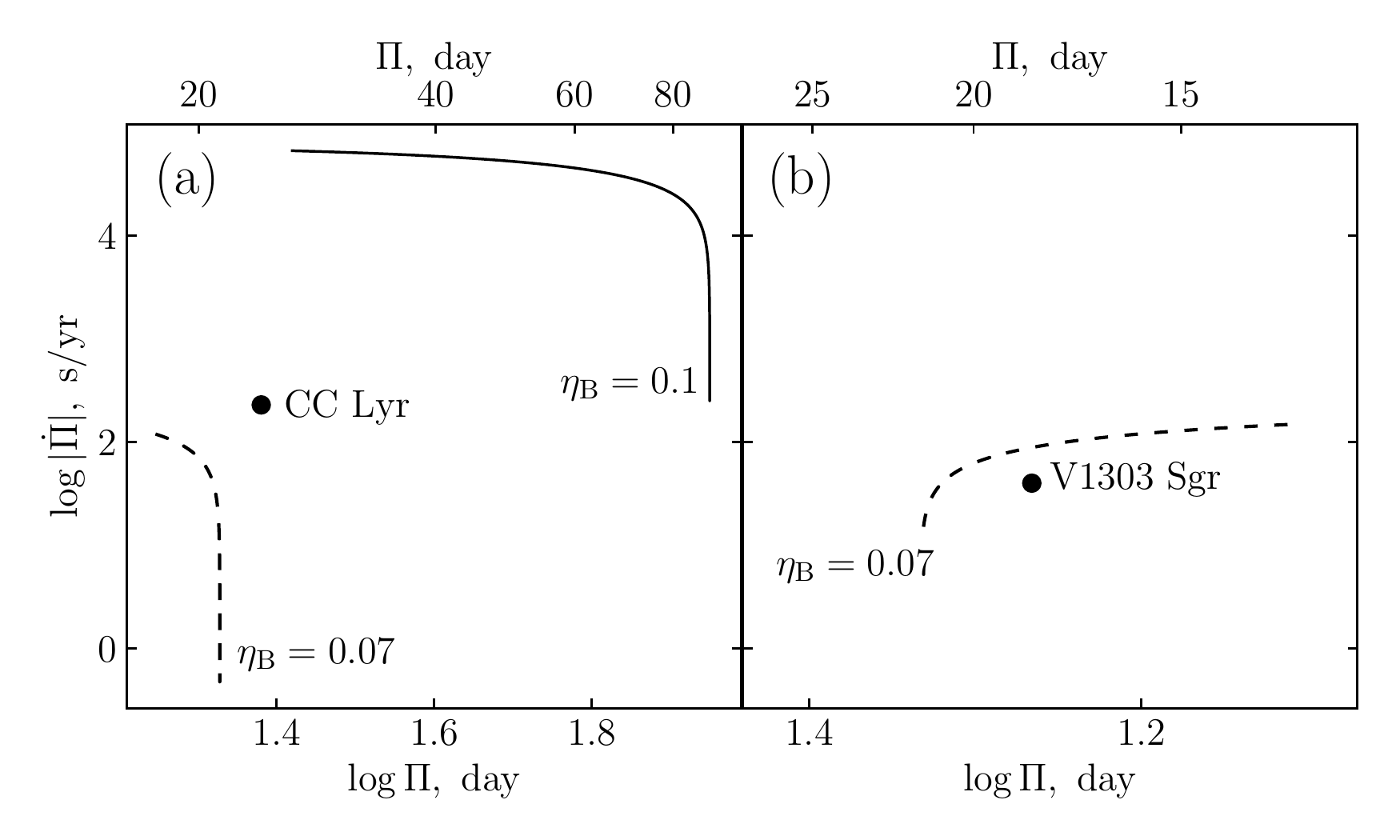}}
\caption{The rate of period change $\dot\Pi$ as a function of period $\Pi$
    after the second crossing of the blue edge of the pulsation instability region
    during the stage of period increase (a) and period decrease (b).
    Evolutionary sequences $\etab=0.07$ and $\etab=0.1$ are shown by solid and dashed lines.
    Observational estimates of $\Pi$ and $\dot\Pi$ for CC~Lyr (Berdnikov et al. 2020)  and
    V1303~Sgr (Barton 1986) are shown by filled circles.}
\label{fig5}
\end{figure}

Variations of $\Pi$ and $\dot\Pi$ between the second crossing of the instability edge and
the maximum period are shown in Fig.~\ref{fig5}a.
In models of the evolutionary sequence $\etab=0.07$ the pulsation period increases from
$\Pi=17.6$ day to $\Pi=21.2$ day during $\approx 5800$ yr and the mass of the hydrogen
envelope ($\Menv=0.011M$) almost does not change.
In hydrodynamic models of the evolutionary sequence $\etab=0.1$ the period of radial
oscillations increases from 28.3 day to 89 day during 165 yr, whereas the mass of
the hydrogen envelope decreases from $\Menv=4.3\times 10^{-3}$ to $\Menv\approx 10^{-3}$.
Thus the W~Vir star with mass $M=0.526M_\odot$ becomes the semiregular pulsating variable
during one and a half century.

Hydrodynamic calculations of radial pulsations for the stage with decreasing period after the
second crossing of the blue edge of the pulsation instability region were done only for
the models of the evolutionary sequence $\etab=0.07$ because for $\etab=0.1$ pulsation periods
significantly exceed those of W~Vir pulsating variables.
Results of these calculations are shown in Fig.~\ref{fig5}b.
The mass of the hydrogen envelope reduces from $\Menv=0.011M$ to $4.6\times 10^{-3}M$
whereas the pulsation period decreases from 21.4 day to 12.7 day during $\approx 9\times 10^3$ yr.

Nonlinearity of radial pulsations in W~Vir variables is responsible for the absence
of strict repetition of light variations which is typical for the most of classical Cepheids.
This feature is a substantial obstacle in determination of the rates of period change so that
only for a few W~Vir stars the successfull measurements of $\dot\Pi$ have been done by now.
To compare results of our calculations with observations we considered two W~Vir pulsating
variables with most reliable $O-C$ diagrams.
One of them is CC~Lyr with period $\Pi=24.01$ day and the rate of period increase
$\dot\Pi =229.3$~s/yr (Berdnikov et al. 2020) and another is V1303~Cyg with period
$\Pi=18.45$~day and the rate of period decrease $\dot\Pi = -40$~s/yr (Barton 1986).
As seen in Fig.~\ref{fig5}, results of our calculations do not contradict
the observations.

\section*{conclusions}

Presented above results of consistent stellar evolution and nonlinear stellar pulsation
calculations allow us conclude that W~Vir pulsating variables are the post--AGB stars
that experience the final helium flash.
Only in this case the luminosity of the post--AGB star is in the range
$10^3L_\odot\lesssim L\lesssim 2\times 10^3L_\odot$,
whereas the periods of radial oscillations
($10~\textrm{day}\lesssim\Pi\lesssim 50~\textrm{day}$)
correspond to periods of light variations of W~Vir stars.
It should be noted that results of our calculations confirm the conclusion of Gingold (1974)
that the helium shell flash before the onset of the post--AGB stage is the necessary
condition for occurence of W~Vir variability.
The term 'final thermal pulse' appeared later in papers by Iben (1982) and Schoenberner (1983).

Evolution of the post--AGB star after the final helium flash proceeds in the thermal timescale
of the hydrogen envelope.
The small envelope mass and large contribution of the radiation field in the internal
energy of the stellar matter are responsible for rapid evolution.
Most demonstrative illustration of rapid evolutionary changes is given by
hydrodynamic models of the evolutionary sequence $\etab=0.1$ after the second crossing
of the pulsation instability edge.
One of such objects with period change in time scale of $\sim 10^2$ yr is V725 Sgr.
In the beginning of the twentieth century this pulsating variable was known as
the population II cepheid but later this star became the semiregular pulsating variable
with period $80~\textrm{day} < \Pi < 100~\textrm{day}$ (Percy 2006).

\newpage
\section*{references}

\begin{enumerate}

\item H.A. Abt, Astrophys. J. Suppl. Ser. \textbf{1}, 63 (1954).

\item A.S. Barton, J. Am. Associat. Var. Star Observ. \textbf{15}, 246 (1986).

\item L.N. Berdnikov, A.M. Yakob, E.N. Pastukhova, Astron. Lett. \textbf{46}, 630 (2020).

\item T. Bl\"ocker, Astron. Astrophys. \textbf{297}, 727 (1995).

\item E. B\"ohm--Vitense, Zeitschrift f\"ur Astrophys. \textbf{46}, 108 (1958).

\item J.R. Buchler and G. Kovacs, Astrophys. J. \textbf{320}, L57 (1987).

\item M. Catelan, Astrophys. J. \textbf{600}, 409 (2004).

\item R.H. Cyburt, A.M. Amthor, R. Ferguson, Z. Meisel, K. Smith,
      S. Warren, A. Heger, R.D. Hoffman, T. Rauscher, A. Sakharuk, H. Schatz,
      F.K. Thielemann, and M. Wiescher,
      Astrophys. J. Suppl. Ser. \textbf{189}, 240 (2010).

\item Yu.A. Fadeyev, Astron. Lett. \textbf{39}, 306 (2013).

\item Yu.A. Fadeyev, Astron. Lett. \textbf{44}, 782 (2018).

\item Yu.A. Fadeyev, Astron. Lett. \textbf{45}, 521 (2019a).

\item Yu.A. Fadeyev, Astron. Lett. \textbf{45}, 655 (2019b).

\item Yu.A. Fadeyev and A.B. Fokin, Astrophys. Space Sci. \textbf{111}, 355 (1985).

\item Yu.A. Fadeyev and H. Muthsam, Astron. Astrophys. \textbf{234}, 188 (1990).

\item R.A. Gingold, Astrophys. J. \textbf{193}, 177 (1974).

\item R.A. Gingold, Astrophys. J. \textbf{204}, 116 (1976).

\item F. Herwig, Astron. Astrophys. \textbf{360}, 952 (2000).

\item I. Iben, \textbf{260}, 821 (1982).

\item A.H. Joy, Astrophys. J. \textbf{86}, 363 (1937).

\item G. Kovacs and J.R. Buchler, Astrophys. J. \textbf{334}, 971 (1988).

\item R. Kuhfu\ss, Astron. Astrophys. \textbf{160}, 116 (1986).

\item M.M. Miller Bertolami, Astron. Astrophys. \textbf{588}, A25 (2016).

\item B. Paxton, R. Smolec, J. Schwab, A. Gautschy, L. Bildsten, M. Cantiello, A. Dotter,
      R. Farmer, J.A. Goldberg, A.S. Jermyn, S.M. Kanbur, P. Marchant, A. Thoul, R.H.D. Townsend,
      W.M. Wolf, M. Zhang, and F.X. Timmes, Astrophys. J. Suppl. Ser. \textbf{243}, 10 (2019).

\item J.R. Percy, A. Molak, H. Lund, D. Overbeek, A.F. Wehlau, and P.F. Williams,
      Publ. Astron, Soc. Pacific \textbf{118}, 805 (2006).

\item D. Reimers, \textit{Problems in stellar atmospheres and envelopes}
      (Ed. B. Baschek, W.H. Kegel, G. Traving, New York: Springer-Verlag, 1975), p. 229.

\item M. Salaris, M. Riello, S. Cassisi, and G. Piotto, Astron. Astrophys. \textbf{420}, 911 (2004).

\item N.N. Samus', E.V. Kazarovets, O.V. Durlevich, N.N. Kireeva, and E.N. Pastukhova,
      Astron. Rep. \textbf{61}, 80 (2017).

\item D. Schoenberner, \textbf{272}, 708 (1983).

\item M. Schwarzschild and R. H\"arm, Astrophys. J. \textbf{160}, 341 (1970).

\item T. Serre, Z. Kollath and J.R. Buchler, Astron. Astrophys. \textbf{311}, 845 (1996).

\item R. Smolec, MNRAS \textbf{456}, 3475 (2016).

\item A.V. Sweigart, J.G. Mengel, and P. Demarque, Astron. Astrophys. \textbf{30}, 13 (1974).

\item R.S. Tuggle and I. Iben, Astrophys. J. \textbf{178}, 455 (1972).

\item G. Wallerstein, Astrophys. J. \textbf{130}, 560 (1959).

\item G. Wallerstein, Publ. Astron, Soc. Pacific \textbf{114}, 689 (2002).

\item G. Wallerstein and A.N. Cox, Publ. Astron, Soc. Pacific \textbf{96}, 677 (1984).

\item C. Whitney, Annales d'Astrophysique \textbf{19}, 142 (1956).

\end{enumerate}

\newpage
\begin{table}
\caption{Main properties of evolutionary sequences}
\label{tabl1}
\begin{center}
\begin{tabular}{c|c|c|c|c|c|c}
\hline
$\etab$ & $n_\mathrm{TP}$ & $\langle\Delta t_\mathrm{ip}\rangle,\ 10^6$ yr& $M/M_\odot$ & $L/L_\odot$ & $\Menv^\star/M_\odot$ & \\
\hline
  0.01 &  7 & 0.234 &  0.550 &  3342 &  0.009 &    \\
  0.02 &  6 & 0.236 &  0.541 &  3093 &  0.005 &    \\
  0.03 &  5 & 0.237 &  0.536 &  1883 &  0.025 &    \\
  0.04 &  5 & 0.238 &  0.535 &  2818 &  0.004 &    \\
  0.05 &  4 & 0.235 &  0.533 &  2637 &  0.062 & $-$\\
  0.06 &  3 & 0.376 &  0.531 &  2291 &  0.033 & $-$\\
  0.07 &  4 & 0.233 &  0.529 &  1866 &  0.035 &    \\
  0.08 &  4 & 0.232 &  0.529 &  5448 &  0.020 & $-$\\
  0.09 &  4 & 0.237 &  0.528 &  2372 &  0.005 &    \\
  0.10 &  4 & 0.236 &  0.528 &  2315 &  0.004 &    \\
  0.11 &  4 & 0.235 &  0.527 &  2218 &  0.004 &    \\
  0.12 &  3 & 0.233 &  0.526 &  2040 &  0.061 & $-$\\
\hline          
 \end{tabular}
\end{center}
\end{table}
\clearpage

\newpage
\clearpage

\newpage
\clearpage

\newpage
\clearpage

\newpage
\clearpage

\newpage
\clearpage

\end{document}